\documentclass[structabstract]{raa}
\usepackage{graphicx,times}
\usepackage{natbib,amssymb}
\usepackage{epsfig}
\usepackage{float}

\providecommand{\bm}[1]{\mbox{\boldmath$#1$\unboldmath}}

\begin{document}
   \title{Accreting CO material onto ONe white dwarfs towards accretion-induced collapse}

   \volnopage{ {\bf 2018} Vol.\ {\bf X} No. {\bf XX}, 000--000}
   \setcounter{page}{1}

   \author{C. Wu \inst{1,2,3,4}
          \and
          B. Wang \inst{1,2,3,4}
          }

   \institute{Yunnan Observatories, Chinese Academy of Sciences, Kunming 650216, China;
          {\it wcy@ynao.ac.cn; wangbo@ynao.ac.cn}\\
          \and
          Key Laboratory for the Structure and Evolution of Celestial Objects, Chinese Academy of Sciences, Kunming 650216, China
          \and
          University of Chinese Academy of Sciences, Beijing 100049, China\\
          \and
          Center for Astronomical Mega-Science, Chinese Academy of Sciences, Beijing 100012, China\\
              }

   \date{Received ; accepted}

\abstract {The final outcomes of accreting ONe white dwarfs (ONe WDs) have been studied for several decades, but there are still some issues not resolved. Recently, some studies suggested that the deflagration of oxygen would occur for accreting ONe WDs with Chandrasekhar masses. In this paper, we aim to investigate whether ONe WDs can experience accretion-induced collapse (AIC) or explosions when their masses approach the Chandrasekhar limit. Employing the stellar evolution code modules for experiments in stellar astrophysics (MESA), we simulate the long-term evolution of ONe WDs by accreting CO material. The ONe WDs undergo weak multicycle carbon flashes during the mass-accretion process, leading to the mass increase of the WDs. We found that different initial WD masses and mass-accretion rates have influence on the evolution of central density and temperature. However, the central temperature cannot reach the explosive oxygen ignition temperature due to the neutrino cooling. This work implies that the final outcome of accreting ONe WDs is electron-capture induced collapse rather than thermonuclear explosion.
\keywords{stars: evolution --- binaries: close --- supernovae: general ---  white dwarfs} }

\titlerunning{Accreting CO material onto ONe WDs towards AIC}

\authorrunning{C. Wu \& B. Wang}

\maketitle


\section{Introduction} \label{1. Introduction}

Carbon-oxygen white dwarfs (CO WDs) in binaries are expected to form type Ia supernovae (SNe Ia) when they grow in mass close to the  Chandrasekar limit ($M_{\rm Ch}$; e.g., Hachisu et al. 1996; Li \& van den Heuvel 1997; Han \& Podsiadlowski 2004; Wang et al. 2009, 2013; Wu et al. 2016; Wang 2018). However, the final outcomes may be changed if the primary stars are oxygen-neon white dwarfs (ONe WDs). The electron-capture induced collapse would occur in an accreting ONe WD when its mass increases to $M_{\rm Ch}$, resulting in the formation of a neutron star (e.g., Miyaji et al. 1980; Nomoto \& Kondo 1991). This process is referred as the accretion-induced collapse (AIC), relating to the formation of some important objects, such as low-mass binary pulsars (LMBPs), millisecond pulsars (MSPs) and magnetars (e.g., Bailyn \& Grindlay 1990; Bhattacharya \& Van den Heuvel 1991; Usov 1992; Dar et al. 1992; Podsiadlowski et al. 2002; Tauris et al. 2013). According to the detailed population synthesis study, Hurley et al. (2010) suggested that the birthrates of binary MSPs from the AIC scenario are comparable to those from the core collapse supernovae, which means that the AIC scenario plays an important role in forming the binary MSPs. AIC may also relate to the r-process nucleosynthesis and the formation of ultrahigh-energy cosmic ray like gamma-ray bursts (e.g., Hartmann et al. 1985; Fryer et al. 1999; Qian \& Wasserburg 2007; Metzger et al. 2008; Piro \& Kollmeier 2016; Lyutikov \& Toonen 2017). In the observations, it is useful in understanding the binary evolution and the formation of neutron stars in globular clusters by identifying the electromagnetic signatures of AIC and constraining its birthrate (e.g., Moriya 2016). Moreover, AIC may also be a potential source of strong gravitational wave emission (e.g., Abdikamalov et al. 2010).

There are two fundamental progenitor models for AIC. (1) The first one is the single-degenerate (SD) model, in which an ONe WD accretes H/He-rich material from its non-degenerate companion via the Roche-lobe overflow. During the mass-transfer process, the accumulated material is transformed into heavier elements due to the multi-shell burning. Eventually, AIC may occur if the ONe WD increases its mass to $M_{\rm Ch}$ (e.g., Nomoto \& Kondo 1991; Langer et al. 2000; Tauris et al. 2013; Brooks et al. 2017). Additionally, CO WD + He star systems may also contribute to the formation of neutron stars through AIC when the off-center carbon ignition occurs on the surface of the WD (e.g., Brooks et al. 2016; Wang et al. 2017). (2) The second one is the double-degenerate (DD) model, in which two WDs in binaries are brought together due to the gravitational wave radiation. The WD will grow in mass via the mass-transfer from its donor. Finally, the primary WD may collapse into a neutron star if the combined mass exceeds $M_{\rm Ch}$ (e.g., Nomoto \& Iben 1985). So far, three merger models have been widely discussed, i.e., the slow merger model, the fast merger model and the composite model. In the slow merger model, the mass ratio of two WDs should be less than $2/3$, and stable mass transfer can occur. The mass donor will be tidally disrupted and form an accretion disc around the primary WD, and the accretion rate from the disc is around the Eddington rate (e.g., Saio \& Jeffery 2000; 2002). In the fast merger model, the merger process is dynamically unstable and may last for a few minutes. All material of the companion will transform onto the surface of the primary and form a corona (e.g., Benz et al. 1990; Schwab et al. 2016). In the composite model, part of the material forms a corona and the remaining material forms a Keplerian disc, in which both the fast and slow merger processes are included (e.g., Yoon et al. 2007; Zhang et al. 2014). Generally, the WD binaries in the DD model could be ONe WD + CO/ONe WD systems (e.g., Schwab et al. 2015; Lyutikov \& Toonen 2017). Some studies also indicated that CO WD + CO WD systems could be progenitors of AIC (e.g., Saio \& Nomoto 2004; Schwab et al. 2016).

However, the final fate of the ONe WD with $M_{\rm Ch}$ has been questioned by some recent studies (e.g., Marquardt et al. 2015; Jones et al. 2016). Since the WD contains nuclear fuel inside (e.g., $^{\rm 16}{\rm O}$, $^{\rm 20}{\rm Ne}$, and maybe $^{\rm 12}{\rm C}$ for less massive WD), and the electron capture as well as the subsequent exothermic reactions of $\gamma$-decay, the ONe WD may undergo explosive oxygen or neon burning and then produce a subpopulation of type Ia supernovae when its mass approaches $M_{\rm Ch}$. Note that whether the ONe WD undergoes collapse or explosion is determined by the competition between electron capture and nuclear burning; the final fate may also be dependent on different conditions of deflagration wave propagation and criteria of explosive nucleosynthesis (e.g., Nomoto \& Kondo 1991; Jones et al. 2016). Therefore, there are still some uncertainties for the final outcomes of ONe WDs with $M_{\rm Ch}$, the study of which is useful in understanding the formation of electron-capture supernovae and type Ia supernovae.

In this paper, we aim to study the long-term evolution of ONe WDs by accreting CO material, and to investigate the final fate of ONe WDs, in which we will consider different initial WD masses (${M}_{\rm WD}^{\rm i}$) and mass-accretion rates ($\dot{M}_{\rm acc}$). In Sect. 2, we introduce our basic assumptions and methods for numerical calculations. The results for our simulations are provided in Sect. 3. Finally, we present discussion and conclusions in Sect. 4.

\section{Methods}

We employ the stellar evolution code named \texttt{Modules for Experiments in Stellar Astrophysics} (MESA; version 7624) (see Paxton et al. 2011, 2013, 2015) to simulate the long-term evolution of the ONe WDs by accreting CO material. Our simulations can be divided into two steps.

Firstly, we use the suite case \texttt{make\_low\_mass\_with\_uniform\_composition} to construct ONe WDs with uniform elemental abundance distribution, in which the mass fractions are $50\%$ $^{\rm 16}{\rm O}$, $45\%$ $^{\rm 20}{\rm Ne}$ and $5\%$ $^{\rm 24}{\rm Mg}$, respectively. These elemental abundances are similar to those of some previous studies (e.g., Takahashi et al. 2013; Schwab et al. 2015). We cool down the ONe cores until their luminosity decreases to $10^{2}\,{L}_\odot$, which we used as our initial WD models.

Secondly, we use the suite case \texttt{wd\_aic} to simulate the mass-accretion processes of the ONe WDs. The nuclear reaction network adopted in our simulations includes $23$ isotopes needed for carbon, oxygen, neon burning (e.g., $^{\rm 12}{\rm C}$, $^{\rm 16}{\rm O}$, $^{\rm 20}{\rm Ne}$, $^{\rm 24}{\rm Mg}$), which are coupled by more than $100$ reactions. We consider the electron-capture reactions for $^{\rm 24}{\rm Mg}$ and $^{\rm 20}{\rm Ne}$ (i.e., $^{\rm 24}{\rm Mg}\,\rightarrow\,^{\rm 24}{\rm Na}\,\rightarrow\,^{\rm 24}{\rm Ne}$; $^{\rm 20}{\rm Ne}\,\rightarrow\,^{\rm 20}{\rm F}\,\rightarrow\,^{\rm 20}{\rm O}$) and $\beta$-decay reactions (the reverse reactions for electron captures). We also consider the ion coulomb corrections and the electron coulomb corrections in our simulations (e.g., Potekhin et al. 2009; Itoh et al. 2002). We use the temperature as the criterion to examine whether AIC can occur. If the central temperature of the WD exceeds a critical value for explosive neon or oxygen burning (e.g., ${\rm log}T_{\rm c}\sim9.6$) at the final stage of evolution, the core will experience thermonuclear explosion supernova, otherwise, the outcome is treated as collapse.

We adopt constant accretion rates to simulate the mass-accretion process, in which the accreted CO material consists of $50\%$ $^{\rm 12}{\rm C}$ and $50\%$ $^{\rm 16}{\rm O}$. The range of accretion rates is $1.0\times10^{-9}\leq\dot{M}_{\rm acc}\leq2.0\times10^{-5}\,{M}_\odot\,\mbox{yr}^{-1}$, which could represent the thick disc model or slow merger model (e.g., Saio \& Jeffery 2000, 2002; Zhang \& Jeffery 2012). In this model, if the mass ratio of two WDs is sufficiently large, the less massive WD in a binary will be dynamically disrupted when it fills its Roche lobe and form a pressure or centrifugalization supported disc surrounding the massive WD. The mass-transfer rate from the pressure supported disc to the primary WD may be close to the Eddington accretion rate. However, the accretion rate could be less than $5.0\times10^{-6}\,{M}_\odot\,\mbox{yr}^{-1}$ for the centrifugalization supported disc and the disc may be last for $10^{4}$ $-$ $10^{6}\,\mbox{yr}$ (e.g., Mochkovitch \& Livio 1990; Yoon et al. 2007; Zhang \& Jeffery 2012; Zhang et al. 2014). This mass-loss mechanism is included in our simulations. Here, we assumed that the super-Eddington wind will be triggered and blows away part of the material from the envelope when the surface luminosity exceeds the Eddington luminosity (see also Denissenkov et al. 2013; Wang et al. 2015; Wu et al. 2017).

\section{Numerical results}

\subsection{An example of AIC}

In Figs\,1-5, we present an example of the long-term evolution of an accreting ONe WD, in which ${M}_{\rm WD}^{\rm i}=1.2{M}_\odot$ and $\dot{M}_{\rm acc}=1.0\times10^{-5}\,{M}_\odot\,\mbox{yr}^{-1}$.

\begin{figure}
\begin{center}
\epsfig{file=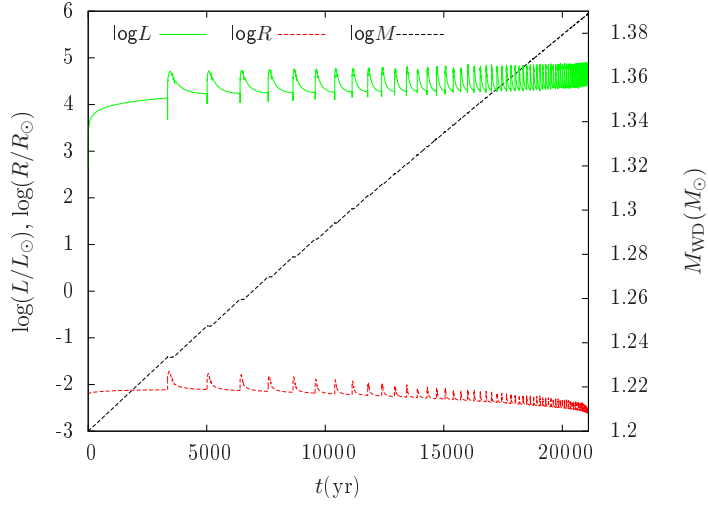,angle=0,width=11.2cm}
 \caption{Long-term evolution of a $1.2{M}_\odot$ accreting WD that undergoes weak carbon flashes. The green solid line, red dashed line and black dashed line represent the luminosity, the WD radius and the WD mass as the function of time during the accretion process, respectively.}
  \end{center}
\end{figure}

Fig.\,1 shows the luminosity, radius and the mass of the WD as the function of time during the mass accumulation process. The CO material piles up onto the surface of the WD during the mass-accretion process, and the CO envelope is heated up by the gravitational potential energy release, leading to the carbon ignition. The flame in the shell transforms the carbon into neon and magnesium, and these heavier elements are mixed into the ONe core by the convection, resulting in the mass increase of the WD. The super-Eddington wind is triggered when the luminosity exceeds the Eddington luminosity that can blow away part of the material. However, the carbon flashes in our simulations are too weak to blow away too much material, resulting in the mass increase of the WD; the radius of the WD decreases gradually since the ONe core becomes massive. Meanwhile, the luminosity changes about an order of magnitude due to the unstable carbon shell burning and fluctuates frequently as the surface temperature increases.

\begin{figure}
\begin{center}
\epsfig{file=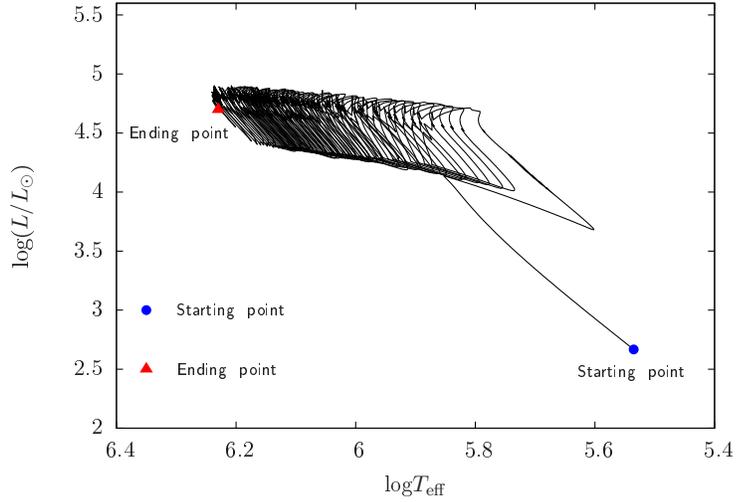,angle=0,width=10.5cm}
 \caption{The Hertzsprung-Russell diagram of $1.2{M}_\odot$ ONe WD during the weak carbon flashes. The blue filled cycle and the red filled triangle represent the starting and ending points, respectively.}
  \end{center}
\end{figure}

Fig.\,2 presents the Hertzsprung-Russell diagram of the accreting WD. The CO material is accumulated onto the surface of the WD at the onset of the accretion process, resulting in the increase of luminosity ($L$) and the effective temperature ($T_{\rm eff}$) via the release of the gravitational potential energy. The carbon in the shell is ignited after $L$ exceeds $10^{4}{L}_\odot$, resulting in the expansion of the envelope and the decrease of $L$ and $T_{\rm eff}$. However, the formed convection zone transfers the heat from the inside out effectively, and the envelope of the WD stops to expand during this stage, leading to the increase of $L$ and $T_{\rm eff}$. The WD gets into the cooling phase rapidly after it experiences the super-Eddington wind stage. The WD experiences more than $60$ carbon flashes during the evolution, and AIC occurs finally when its mass approaches $1.39{M}_\odot$.

\begin{figure}
\begin{center}
\epsfig{file=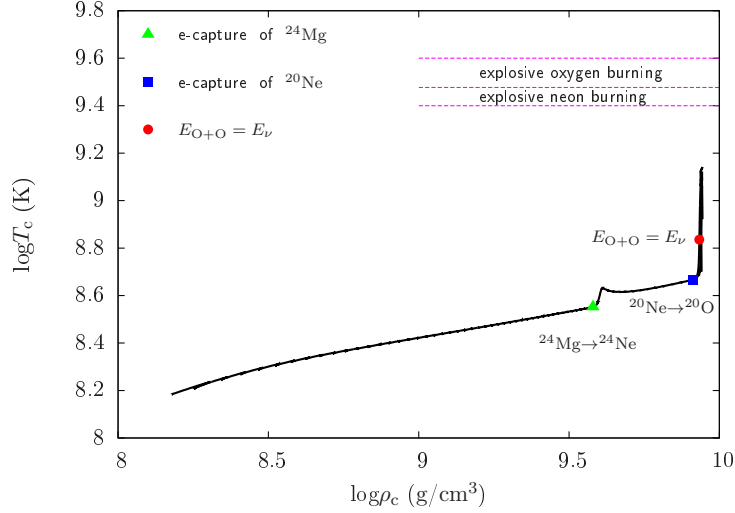,angle=0,width=10.5cm}
 \caption{Central density-temperature profile for the evolution of $1.2{M}_\odot$ ONe WD during mass accumulation. The green triangle and blue square are the starting points of electron-capture reactions of $^{\rm 24}{\rm Mg}$ and $^{\rm 20}{\rm Ne}$, respectively. The red cycle is the point where the nuclear reaction energy generation from oxygen burning equals to the neutrino losses. The pink dashed line shows the critical temperature range for explosive neon and oxygen burning.}
  \end{center}
\end{figure}

Fig.\,3 presents the central density-temperature profile of the ONe WD during the evolution process. The WD becomes massive due to the multiple carbon shell flashes, leading to the increase of the central density ($\rho_{\rm c}$) and temperature ($T_{\rm c}$). After $\rho_{\rm c}$ approaches the point where the electron-capture timescale of $^{\rm 24}{\rm Mg}$ equals to the core compression timescale (i.e., $t_{\rm compress}\,\approx\,5\times10^{4}\,\mbox{yr}\,(\frac{\rho_{\rm c}}{10^{9}\,\mbox{g}\,\mbox{cm}^{-3}})^{-0.55}\,(\frac{\dot{M}}{10^{-6}\,{M}_\odot\,\mbox{yr}^{-1}})^{-1}$; see Schwab et al. 2015), $T_{\rm c}$ increases owing to the exothermic reactions of electron capture until the exhaustion of central $^{\rm 24}{\rm Mg}$. The similar phenomenon of rapid increase of $T_{\rm c}$ occurs when $\rho_{\rm c}$ approaches the blue point, where the electron-capture reaction of $^{\rm 20}{\rm Ne}$ is triggered. Since the mass fraction of $^{\rm 20}{\rm Ne}$ is $45\%$, the electron-capture process of $^{\rm 20}{\rm Ne}$ can last for a long time, resulting in the occurrence of $^{\rm 16}{\rm O} $+$ ^{\rm 16}{\rm O}$ nuclear reactions. After the oxygen burning is ignited, $T_{\rm c}$ increases obviously owing to the exothermic process through the electron capture and oxygen burning. However, the neutrino energy loss rate ${\varepsilon}_{\nu}\propto{T}^{9}$ when the temperature exceeds $10^{9}{\rm K}$. Subsequently, the energy release of the nuclear reactions cannot balance the energy loss at the high temperature, leading to the decrease of ${T}_{\rm c}$. The highest $T_{\rm c}$ of the WD in our simulations is $1.58\times10^{9}{\rm K}$, which is lower than $3\times10^{9}{\rm K}$; the explosive oxygen burning can occur if the temperature exceeds this value. This implies that the oxygen explosion cannot occur and the WD will form a neutron star through AIC eventually.

\begin{figure}
\begin{center}
\epsfig{file=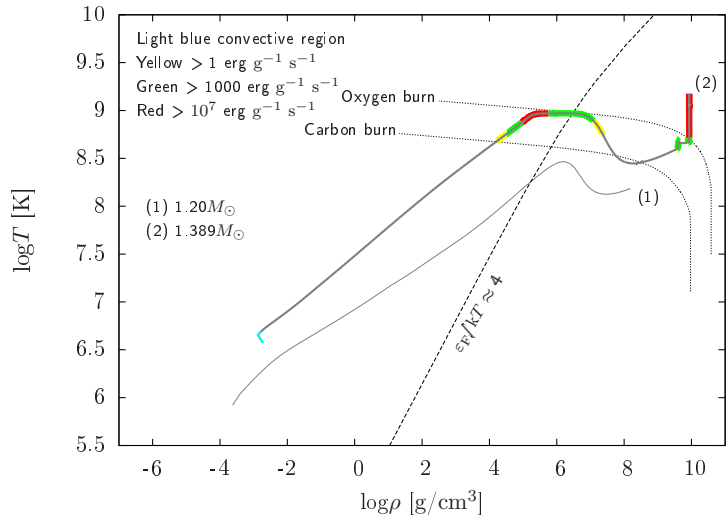,angle=0,width=10.5cm}
 \caption{Density-temperature profile of the initial and final ONe WD. The initial profile is represented by line (1), whereas the final profile is represented by line (2).}
  \end{center}
\end{figure}

Fig.\,4 shows the density-temperature profiles of the ONe WD in the initial and final stage. At the onset of the accretion process, the surface of the WD is compressed by the material, leading to the formation of the convection zone on the surface. The $^{\rm 12}{\rm C} $+$ ^{\rm 12}{\rm C}$ nuclear reactions are triggered when the ignition conditions of the carbon shell are satisfied, transforming the carbon into heavier elements (e.g., $^{\rm 20}{\rm Ne}$ and $^{\rm 24}{\rm Mg}$). The ONe WD grows in mass continuously and reaches $1.389{M}_\odot$ eventually. Note that there are two burning regions in the center of the core in the final stage. The burning zone for the relatively low density position (${\rm log}{\rho}_{\rm c}\sim9.6$) is caused by the electron-capture reactions of $^{\rm 24}{\rm Mg}$, whereas the innermost zone is caused by the electron-capture reactions of $^{\rm 20}{\rm Ne}$ and oxygen burning, in which the central nuclear reaction rate exceeds $2.5\times10^{8}\,\mbox{erg}\,\mbox{g}^{-1}\,\mbox{s}^{-1}$.

\begin{figure}
\begin{center}
\epsfig{file=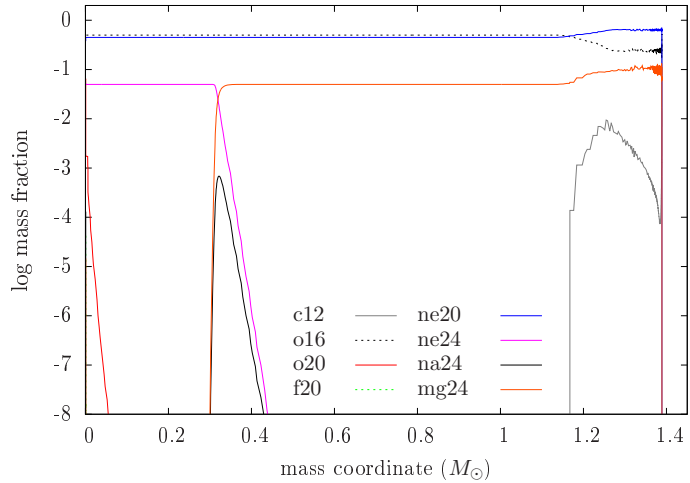,angle=0,width=10.5cm}
 \caption{Elemental abundance distribution profile of the ONe WD at the final moment.}
  \end{center}
\end{figure}

Fig.\,5 presents the elemental abundance distribution of the ONe WD in the final stage, which can provide initial input parameters in simulating the AIC processes. The nuclear fusion transforms the carbon into $^{\rm 20}{\rm Ne}$ and $^{\rm 24}{\rm Mg}$ after the flashes occur, resulting in a higher abundances of neon and magnesium on the surface. In the center of the WD, the exothermal electron-capture reactions of $^{\rm 24}{\rm Mg}$ generate a burning wave, which propagates outside the core and transforms all $^{\rm 24}{\rm Mg}$ into $^{\rm 24}{\rm Ne}$ in the center. The abundance of $^{\rm 20}{\rm O}$ in the center increases sharply after the electron-capture reaction chain $^{\rm 20}{\rm Ne}\,\rightarrow\,^{\rm 20}{\rm F}\,\rightarrow\,^{\rm 20}{\rm O}$ is triggered. However, the temperature and density needed for electron-capture reaction of $^{\rm 20}{\rm F}$ is much lower than that of $^{\rm 20}{\rm Ne}$, resulting in that the $^{\rm 20}{\rm F}$ is transformed into $^{\rm 20}{\rm O}$ immediately once it appears (see Schwab et al. 2015).

\subsection{Results of different initial WD masses and accretion rates}

\begin{table*}
\centering
\caption{Parameters of accreting ONe WDs with different initial masses and accretion rates. ${M}_{\rm WD}^{\rm i}$ = initial WD mass; $\dot{M}_{\rm acc}$ = mass accretion rate; $t_{\rm e}$ = evolution time; ${M}_{\rm WD}^{\rm f}$ = final WD mass; ${\rm log}{R}_{\rm WD}^{\rm f}$ = final WD radius; ${\rm log}{\rho}_{\rm c}^{\rm f}$ = final central density of the WD; ${\rm log}{T}_{\rm c}^{\rm f}$ = final central temperature of the WD.}
\begin{tabular}{|c|c|c|c|c|c|c|}     
\hline
 ${M}_{\rm WD}^{\rm i} ({M}_\odot)$ & $\dot{M}_{\rm acc} ({M}_\odot\,\mbox{yr}^{-1})$ & $t_{\rm e} (\mbox{yr})$ & ${M}_{\rm WD}^{\rm f} ({M}_\odot)$ & ${\rm log}{R}_{\rm WD}^{\rm f} ({R}_\odot)$ & ${\rm log}{\rho}_{\rm c}^{\rm f} (\mbox{g}\,\mbox{cm}^{-3})$ & ${\rm log}{T}_{\rm c}^{\rm f} ({\rm K})$\\
 \hline
 1.20 & $2.0\times10^{-5}$ & $1.323\times10^{4}$ & $1.389$ & -2.532 & 9.9386 & 8.8892\\
 \hline
 1.15 & $1.0\times10^{-5}$ & $3.259\times10^{4}$ & $1.389$ & -2.624 & 9.9366 & 8.9943\\
 1.20 & $1.0\times10^{-5}$ & $2.111\times10^{4}$ & $1.389$ & -2.572 & 9.9394 & 8.7058\\
 1.30 & $1.0\times10^{-5}$ & $9.851\times10^{3}$ & $1.389$ & -2.633 & 9.9744 & 8.8057\\
 \hline
 1.20 & $1.0\times10^{-6}$ & $1.873\times10^{5}$ & $1.387$ & -2.683 & 9.9407 & 9.1180\\
 1.20 & $1.0\times10^{-7}$ & $1.859\times10^{6}$ & $1.386$ & -2.707 & 9.9514 & 9.0288\\
 1.20 & $1.0\times10^{-8}$ & $1.840\times10^{7}$ & $1.384$ & -2.708 & 9.9664 & 9.1528\\
 1.20 & $1.0\times10^{-9}$ & $1.830\times10^{8}$ & $1.383$ & -2.714 & 9.9811 & 8.8240\\
  \hline
\end{tabular}
\end{table*}

In order to investigate whether the initial WD mass (${M}_{\rm WD}^{\rm i}$) and accretion rate ($\dot{M}_{\rm acc}$) have influence on the final fate of accreting ONe WDs or not, we simulate the long-term evolution of accreting ONe WDs with different ${M}_{\rm WD}^{\rm i}$ and $\dot{M}_{\rm acc}$. The results are summarized in Table\,1.

\begin{figure}
\begin{center}
\epsfig{file=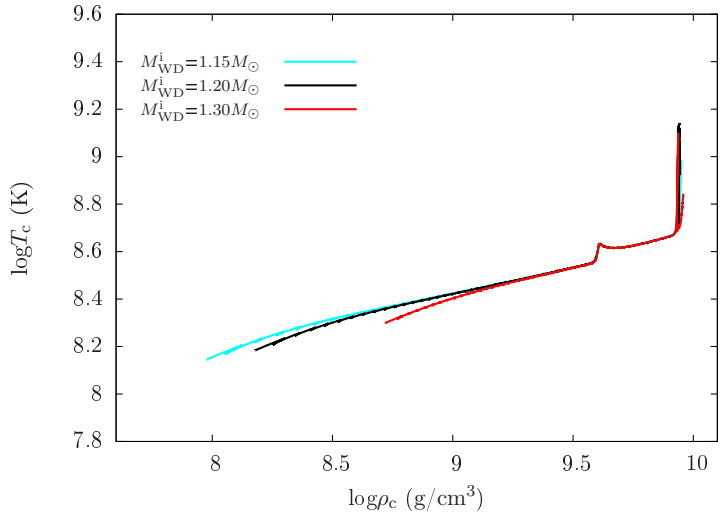,angle=0,width=10.5cm}
 \caption{Central density-temperature profiles for the evolution of accreting ONe WDs, in which $\dot{M}_{\rm acc}=1.0\times10^{-5}\,{M}_\odot\,\mbox{yr}^{-1}$.}
  \end{center}
\end{figure}

Fig.\,6 presents the central density-temperature evolutionary profiles for the different initial masses of accreting ONe WDs, in which ${M}_{\rm WD}^{\rm i}$=1.15, 1.2 and $1.3{M}_\odot$. In this simulation, $\dot{M}_{\rm acc}=1.0\times10^{-5}\,{M}_\odot\,\mbox{yr}^{-1}$. Although ${\rho}_{\rm c}$ and $T_{\rm c}$ are different at the onset of the simulations, they are tend to be similar as the WDs increase in mass. The electron-capture reactions of $^{\rm 24}{\rm Mg}$ and $^{\rm 20}{\rm Ne}$ occur at the same temperature and density, and all of the WDs experience AIC at the end of calculations. This implies that different initial WD masses have negligible influence on the final outcome of accreting ONe WDs. Note that a bifurcation appears at the end of the evolutionary track of $1.3{M}_\odot$ due to numerical reasons, which has no influence on the final outcome.

\begin{figure}
\begin{center}
\epsfig{file=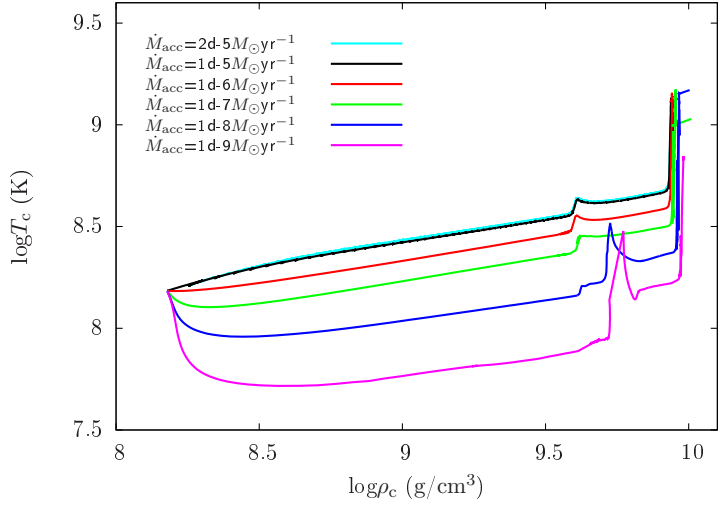,angle=0,width=10.5cm}
 \caption{Similar to Fig.\,6, but for various accretion rates, in which ${M}_{\rm WD}^{\rm i}=1.2{M}_\odot$.}
   \end{center}
\end{figure}

The accretion rates may also have influence on the mass-accretion timescale and the corresponding evolution of $T_{\rm c}$. Here we simulate the evolution of $1.2{M}_\odot$ ONe WD with a wide range of $\dot{M}_{\rm acc}$. In Fig.\,7, we present the central density-temperature profiles of the accreting ONe WD with different $\dot{M}_{\rm acc}$. For low $\dot{M}_{\rm acc}$, $T_{\rm c}$ is lower due to longer evolution timescale of the accreting WDs. Note that $\dot{M}_{\rm acc}$ has little influence on $\rho_{\rm c}$ at the onset of electron-capture reactions of $^{\rm 20}{\rm Ne}$. However, the profiles are not similar for that of $^{\rm 24}{\rm Mg}$, especially in the cases of $\dot{M}_{\rm acc}=1.0\times10^{-8}$ and $1.0\times10^{-9}\,{M}_\odot\,\mbox{yr}^{-1}$. This is because for lower $\dot{M}_{\rm acc}$ the electron-capture reaction $^{\rm 24}{\rm Mg}\,\rightarrow\,^{\rm 24}{\rm Na}$ occurs at a relative lower temperature, but the electron-capture reaction of $^{\rm 24}{\rm Na}$ is not be triggered (see Fig.\,4 in Schwab et al. 2015). The conditions in the central zone will be satisfied with the reaction chain $^{\rm 24}{\rm Na}\,\rightarrow\,^{\rm 24}{\rm Ne}$ as $\rho_{\rm c}$ increases, leading to the subsequent increase of $T_{\rm c}$. Thus, the evolutionary tracks of the density-temperature are different for $\dot{M}_{\rm acc}=1.0\times10^{-8}$ and $1.0\times10^{-9}\,{M}_\odot\,\mbox{yr}^{-1}$.

\begin{figure}
\begin{center}
\epsfig{file=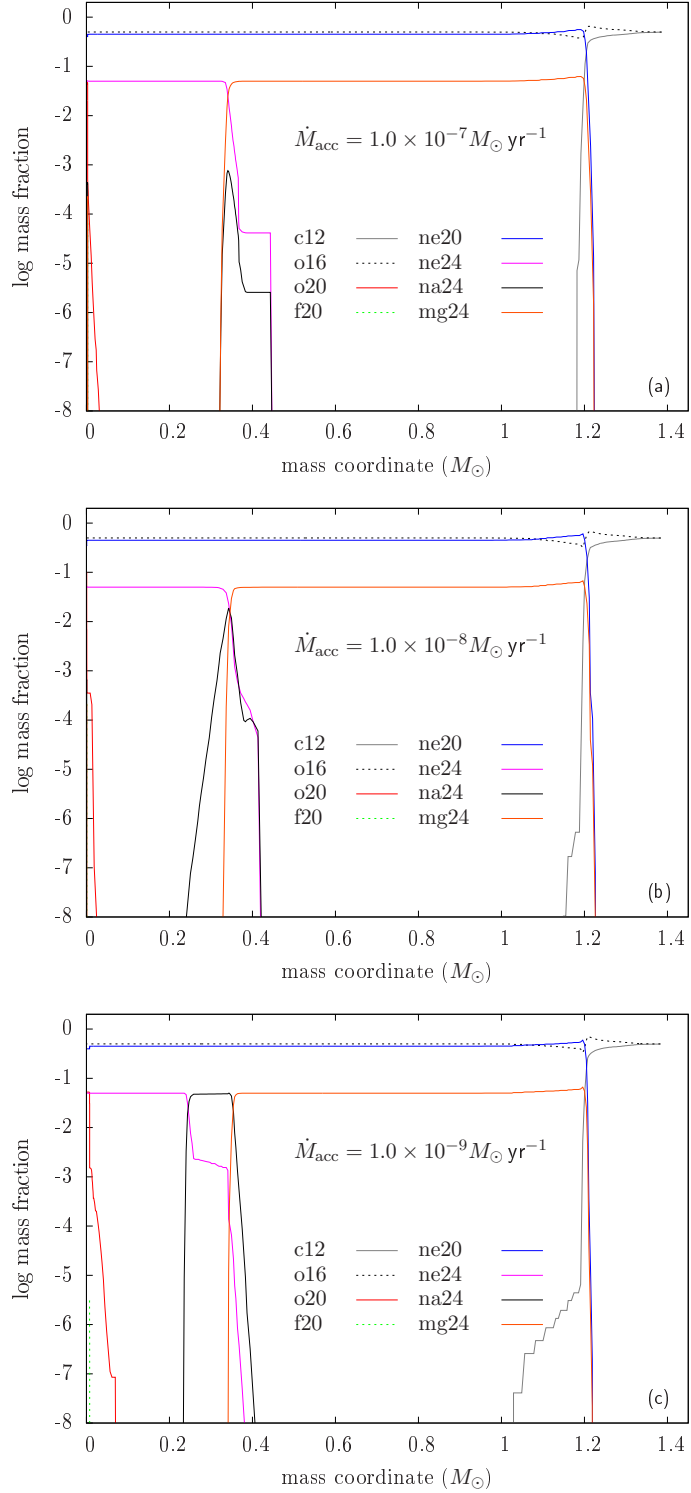,angle=0,width=10.5cm}
 \caption{Similar to Fig.\,5, but for various accretion rates, in which ${M}_{\rm WD}^{\rm i}=1.2{M}_\odot$. Panel (a): $\dot{M}_{\rm acc}=1.0\times10^{-7}\,{M}_\odot\,\mbox{yr}^{-1}$; Panel (b): $\dot{M}_{\rm acc}=1.0\times10^{-8}\,{M}_\odot\,\mbox{yr}^{-1}$; Panel (c): $\dot{M}_{\rm acc}=1.0\times10^{-9}\,{M}_\odot\,\mbox{yr}^{-1}$.}
   \end{center}
\end{figure}

In Fig.\,8, we present the elemental abundance distribution profiles in the final stage of $1.2{M}_\odot$ ONe WD with three different accretion rates. Note that the elemental abundances of carbon in the shells are different with that of Fig.\,5. This is because for the low $\dot{M}_{\rm acc}$ the surface temperatures are lower due to longer evolution timescale, resulting in that the carbon burning does not occur in the shell of the WDs. Meanwhile, for lower $\dot{M}_{\rm acc}$ the temperature of the core is cooler when the electron-capture reaction of $^{\rm 24}{\rm Mg}$ occurs. At the relatively low temperature, the electron-capture rate of $^{\rm 24}{\rm Mg}$ is higher than that of $^{\rm 24}{\rm Na}$. Thus, the elemental abundance of $^{\rm 24}{\rm Na}$ in the core becomes higher as $\dot{M}_{\rm acc}$ decreases. Similarly, the occurrence of electron-capture reactions of $^{\rm 20}{\rm Ne}$ needs higher $\rho_{\rm c}$ when core is cooler (see Fig.\,4 in Schwab et al. 2015). This indicates that the core becomes denser for the lower $\dot{M}_{\rm acc}$ when AIC occurs (see also Fig.\,5). Thus, the final WD radiuses in Table\,1 are smaller for lower $\dot{M}_{\rm acc}$.

\section{Discussion and Conclusions} \label{6. DISCUSSION}

The WDs in a binary would merge together due to the loss of orbital angular momentum. However, the merger process is still under debate. Several studies suggested that the fast merger may occur if the mass ratio of two WDs is larger than 2/3, in which the WD binary would merge dynamically, resulting in a high mass-transfer rate (e.g., $10^{4}\,{M}_\odot\,\mbox{yr}^{-1}$; see Zhang \& Jeffery 2012). Yoon et al. (2007) indicated that not all of the material from the secondary transfers onto the primary WD to form a hot corona. In their model, the remaining material could form a Keplerian disc surrounding the primary WD, and the mass-transfer rate from the Keplerian disc may be less than $5.0\times10^{-6}\,{M}_\odot\,\mbox{yr}^{-1}$. Note that the merger models still remain uncertain, especially when rotation is taken into account; Saio \& Nomoto (2004) suggested that the accretion rates may decrease if the primary WD approaches the critical rotation. In our simulations, we only considered the constant accretion rates to represent the thick disc model. The stable mass transfer would occur during the accretion, and the mass-accretion rate from the thick disc is usually less than $2.0\times10^{-5}\,{M}_\odot\,\mbox{yr}^{-1}$. This indicates that our assumption is feasible.

The final outcomes of ONe core have been investigated by some studies so far. However, it is still not understood completely. Since the ONe core contains nuclear fuel inside, the WD may undergo oxygen or neon deflagration when its mass approaches $M_{\rm Ch}$. Isern et al. (1991) found that a thermonuclear supernova may be produced if the ignition density of oxygen is lower than $9.5\times10^{9}\,\mbox{g}\,\mbox{cm}^{-3}$. Recently, Jones et al. (2016) simulated the oxygen deflagration processes of the degenerate ONe core. Their results supported the 1D simulations of Isern et al. (1991), and also suggested that the ONe WD can collapse into neutron star if the ignition density of oxygen is relatively high (e.g., $2\times10^{10}\,\mbox{g}\,\mbox{cm}^{-3}$). In our simulations, we used the temperature as the criterion for AIC, which means that the AIC can be triggered if ${\rm log}T_{\rm c}$ of the ONe WD during the final stage cannot exceed $9.6$ (i.e., the temperature of explosive oxygen ignition). The temperature criterion is also supported by some studies (e.g., Schwab et al. 2015). Our simulations shows that $T_{\rm c}$ cannot exceed $1.58\times10^{9}{\rm K}$ due to the neutrino-loss mechanism. This implies that ${\rho}_{\rm c}$ will increase until AIC occurs.

In our simulations, we set the initial $X_{\rm Mg}$ to be $5\%$. However, various $X_{\rm Mg}$ may have influence on the evolution of $T_{\rm c}$ and ${\rho}_{\rm c}$ when electron-capture reaction of $^{\rm 24}{\rm Mg}$ is triggered. Schwab et al. (2015) explored the evolutions of ONe core with a wide range of $X_{\rm Mg}$ (i.e., 0.01 - 0.2). In their studies, $T_{\rm c}$ and ${\rho}_{\rm c}$ of the ONe WDs have the similar evolution tracks if $X_{\rm Mg}\leq0.05$, but for higher $X_{\rm Mg}$ the electron-capture reaction of $^{\rm 20}{\rm Ne}$ is triggered in a relative high ${\rho}_{\rm c}$. This implies that $X_{\rm Mg}$ in the initial WDs may not affect the final outcome of ONe cores (i.e., collapse into a neutron star) when their masses approach $M_{\rm Ch}$ based on the results of Jones et al. (2016). Meanwhile, we have not considered the Urca-process cooling in our simulations, which may have influence on the evolution of accreting ONe WD. However, Schwab et al. (2017) found that ${\rho}_{\rm c}$ of the ONe WD is a little bit larger than that without Urca-process cooling when oxygen is ignited, which means that the final outcome of ONe cores will not be changed.

Comparing with the standard core collapse channel, NSs originated from AIC channel may have some special properties. (1) AIC may produce NSs with high magnetic fields and slow rotation rates if the AIC process occurs near the end of mass transfer stage, which may explain the existence of NSs with high magnetic fields and slow rotation rates in the close binary systems of Galactic disk (e.g., Yungelson et al. 2002; Tauris et al. 2013). (2) NSs from this channel generally received smaller kick than that through the standard core collapse channel, and the evolution time of progenitor systems for AIC is longer, which may explain the existence of young NSs living in globular clusters (e.g., Boyles et al. 2011; Tauris et al. 2013). (3) The mass donors of NSs from this channel may have ultra-low masses, since a large amount of material are transformed to the primaries during the binary evolution phases (e.g., Lyne et al. 1993).

By employing MESA, we investigated the long-term evolution of ONe WDs during the mass-accretion process. We adopted different initial WD masses (i.e., ${M}_{\rm WD}^{\rm i}$=1.15, 1.2 and $1.3{M}_\odot$) and accretion rates (i.e., $1.0\times10^{-9}\leq\dot{M}_{\rm acc}\leq2.0\times10^{-5}\,{M}_\odot\,\mbox{yr}^{-1}$) in our simulations. Our results indicate that ${M}_{\rm WD}^{\rm i}$ and $\dot{M}_{\rm acc}$ could affect the evolution of $T_{\rm c}$ and ${\rho}_{\rm c}$ of the WDs, but the explosive oxygen ignition cannot be triggered by the exothermic process of electron-capture reactions. This implies that ONe WDs will finally undergo AIC to form neutron stars. Note that rotation may have influence on the structure of the accretion disc and the subsequent outcomes of the ONe WDs (e.g., Yoon et al. 2007). However, the current study on the influence of rotation still remains uncertain, which needs to be investigated further. Finally, we hope that AIC events can be identified in the future, which will be useful for the studies of binary evolution and electron-capture supernovae.

\begin{acknowledgements}
We acknowledge useful comments and suggestions from the referee.
This study is supported by the National Basic Research Program of China (973 programme, 2014CB845700),
the Chinese Academy of Sciences (Nos KJZD-EW-M06-01 and QYZDB-SSW-SYS001),
the National Natural Science Foundation of China (Nos 11673059, 11521303, 11390374 and 11573016),
and the Natural Science Foundation of Yunnan Province (Nos 2013HB097, 2013HA005 and 2017HC018).

\end{acknowledgements}

\label{lastpage}

\end{document}